# On the Complexity of Maximum Clique Algorithms: usage of coloring heuristics leads to the $\Omega(2^{0.2n})$ algorithm running time lower bound.

Lavnikevich Nikolay,

*NET1 Universal Technologies (Vostok)*

**Abstract.**

Maximum Clique Problem(MCP) is one of the 21 original NP -complete problems enumerated by Karp[19] in 1972. In the last years a large number of exact methods to solve MCP have been appeared[2, 3, 6, 8, 9, 13, 14,17,18,21,23,25,29-31,34-38,40-43]. Most of them are branch and bound algorithms and use branching rule introduced by Balas and Yu[3] and based on coloring heuristics to establish an upper bound on the clique number. They differ from each other primarily in vertex preordering and vertex coloring methods.

Current methods of worst case running time analysis for branch and bound algorithms do not allow to provide tight upper bounds. This motivates the study of lower bounds for such algorithms. We prove $\Omega(2^{0.2n})$ lower bound for group of MCP algorithms based on usage of coloring heuristics.

# 1. Introduction

A *computational problem* consists in the following: for a given input(taken from a domain of instances) compute a solution, that is an output which satisfies a given relation with the input. For example the sorting problem consists in computing, for a given list of objects (say, integers), a permutation of the objects in non-decreasing order (according to a given order relation).

An *algorithm* is a well-defined step-by-step computational procedure which takes some inputs and provides some outputs. It can be viewed as a program which runs on a given machine. An algorithm solves a computational problem if, for every given instance of the problem, it computes a (correct) solution in a finite number of steps.

When more than one algorithm is available to solve a given problem, we often prefer to use the most efficient one, that is the algorithm which uses the smallest amount of a given resource. The resource considered is usually *time*, measured as the number of steps required to solve the problem (what can be done in one step, depends on the machine model considered) or *memory*.

Since the time an algorithm takes to solve a problem may be different for each instance considered, we need a simple and easy qualitative way to summarize it. The *(worst-case) time complexity* of an algorithm is the maximum number of steps required to solve an instance of *order n*.

Determining the exact time complexity of an algorithm can be difficult, tedious and not really relevant (most of the times, all we need is its order of magnitude). For these reasons, the *asymptotic time complexity* is often preferred.

Suppose $f : Z \to R$ and $g : Z \to R$ are functions.

We say f is $O(g)$ if there exists constants $c_1$ and $n_1$ so that $|f(n)| \leq c_1*|g(n)|$ for all $n > n_1$. In other words, f is $O(g)$ if it is never larger than a constant times g for all large values of n. The function $c_1*g(n)$ gives an *asymptotic upper bound* on the size of f(n) for all values of $n > n_1$.

Usually the expression for g is less complex then expression for f, and that is one of the things that makes big-O notation useful. Notice that we don't care what happens for "small" values of n. Also, usually we don't worry too much about the absolute value signs since we usually compare functions that take positive values.

We say f is $\Omega(g)$ if there exists constants $c_2$ and $n_2$ so that $|f(n)| \geq c_2*|g(n)|$ for all $n > n_2$. Big-$\Omega$ is just like big-O, except that $c_2*g(n)$ is an *asymptotic lower bound* for f(n) for all values of $n > n_2$.

Eventually, an algorithm is $\Theta(g)$ if it is O(g) and $\Omega(g)$.

A *polynomial-time algorithm* is an algorithm whose time complexity is *O(p(n))*, for some polynomial *p(n)* of n.

All the other algorithms are usually referred as *exponential-time algorithms.* The distinction between polynomial-time and exponential-time algorithms provides a simple way to separate *tractable* problems (that is, solvable in polynomial-time) from *intractable* or $NP-complete$(1971, Cook[10]) one.

*Maximum Clique* Problem is $NP-$complete[19].

## 2. Notations and Definitions

Given an arbitrary *simple undirected graph G = (V, E)* without loops and multiple edges. The *order* of a graph is the number of vertices $n=|V|$. A graph's *size* is the number of edges $m=|E|$.

A *complement* of graph *G* is the graph $\overline{G}$ = (V, $\overline{E}$) where $\overline{E}$ = { *(v, u)* | v, u $\in$ V, v $\neq$ u, and *(v, u)* $\notin$ E }.

A *clique Q* is a set of pair wise adjacent vertices of the graph. The *maximum clique problem*(MCP) is to find a clique of maximum cardinality in a graph *G*.

An *independent set (stable set, vertex packing) I* is a set of pair wise nonadjacent vertices of the graph. The *maximum independent set* (MIS) problem is to find an independent set of maximum cardinality in a graph *G*.

A *vertex cover C* is a subset of V such that every edge *(v, u)* $\in$ E is incident to at least one vertex in S. The *minimum vertex cover* (MVC) problem is to find a vertex cover of minimum cardinality in a graph *G*.

It is easy to see that *Q* is a clique in a graph *G = (V, E)* if and only if *V - Q* is a vertex cover in the complement graph $\overline{G}$ = *(V, $\overline{E}$)* , and if and only if *Q* is an independent set of $\overline{G}$. Thus, the maximum clique problem, the vertex cover problem and the maximum independent set problem are equivalent.

In addition, they are all $\mathcal{NP}$—complete, which means that unless $\mathcal{P}= \mathcal{NP}$ there exists no algorithm that can solve this problems in time polynomial to the order of the input graph.

For $k≥1$ a *k-coloring* of G is a mapping $\varphi$ of $V(G)$ into the (color-) set $\{1, …, k\}$ satisfying $\varphi(v) ≠ \varphi(u)$ for any adjacent vertices $v, u \in V$. A graph which admits a *k-coloring* is called *k-colorable*.

The *chromatic number* $\chi(G)$ of a non-empty graph G is the smallest integer $k$ for which G is *k*-colorable. If $\chi(G) = k$ then G is called *k-chromatic*.

Note that $H \subseteq G$ implies $\chi(H) \leq \chi(G)$.

**Proposition 2.1** *For every graph G the clique number w(G) is a lower bound on the chromatic number $\chi$(G)*

$$w(G) \leq \chi(G).$$

The difference between the chromatic number and the clique number of a graph Gyárfás [16] proposed to call the *chromatic gap*. We denote this value by

$$\psi(G) = \chi(G) - w(G)$$

## 3. Maximum Clique Algorithm Complexity

A lot of effort has been devoted in last decades to develop faster and faster exact algorithms MC problem. Let us remember the main achieved results.

$O(2^n)$. Zykov(1962, p.208) [43] proposed classical graph decomposition schema $L \rightarrow [L_p, L_r]$ :

"*Let the graph L vertices are numbered and let $L_p$ – subgraph induced by all vertices adjacent with the first vertex; $L_r$ –subgraph induced by all vertices other than the first.*"

This decomposition schema causes classical $O(2^n)$-time recursive algorithm

$$max \{ 1 + w(N(v)), w(G – \{v\}) \}.$$

$\Theta(2^{0.528n})$. By a result of Moon & Moser (1965) [24], any *n*-vertex graph has at most $3^{n/3}$ maximal cliques. Therefore, any algorithm which enumerates all maximal cliques of a graph on n vertices must have worst case running time of $\Omega(3^{n/3})$. The Bron–Kerbosch(1973) [7] algorithm is a recursive backtracking procedure that augments a candidate clique by considering one vertex at a time, either adding it to the candidate clique or to a set of excluded vertices that cannot be in the clique but must have some non-neighbor in the eventual clique.

Tomita[36] presented a depth-first search algorithm for generating all maximal cliques of an undirected graph, in which pruning methods are employed as in the Bron–Kerbosch algorithm and proved its worst-case time complexity $O(3^{n/3})$ or $O(2^{0.528n})$.

$O(2^{n/3})$. On the other hand, finding the maximum clique of a graph does not require to actually examine all of its maximal cliques. Along the search among the maximal cliques of the graph, some non-maximal cliques can be discarded as soon as they are identified as not contained in a clique larger than another already known. Tarjan and Trojanowski(1976) [34] proposed such algorithm with worst case running time of $O(2^{n/3})$.

$O(2^{0.304n})$. Jian(1986) [16] improved algorithm complexity to $O(2^{0.304n})$.

$O(2^{0.296n})$ and $O(2^{0.276n})$. In 1986 Robson [30] proposed two versions of a new but similar algorithm: the first version runs in polynomial space in time $O(2^{0.296n})$ and the second version, which used exponential space, needs $O(2^{0.276n})$.

$O(2^{n/4})$. In 2001 Robson[31] reduced this bound to $O(2^{n/4})$ and this bound remaining the best known.

$\Omega(2^{0.166n})$ and $O(2^{0.288n})$. Fomin(2006) [14] proposed $O(2^{0.288n})$ algorithm and simultaneously he showed $\Omega(2^{0.166n})$ lower bound on the worst-case time complexity of his MIS algorithm.

But unfortunately no experimental results/tests of this algorithm are known.

Group of "practical" algorithms commonly use the branch and bound method. The key issues in a branch and bound algorithm for the MCP are Bomze[6]:

1. How to find a good lower bound, i.e. a clique of large order?
2. How to find a good upper bound on the order of maximum clique?
3. How to branch, i.e. break a problem into smaller subproblems?

One of the most important contributions in the 1980's on practical algorithms for the MCP is due to Balas and Yu[3]. They proposed to use on the second phase(upper bound) a well known fact that the chromatic number of a graph is always bigger or equal to the order of this graph clique number

$$w(G) \leq \chi(G).$$

Later appeared a lot of algorithms using different heuristic vertex-coloring on the second phase (Babel[1], Wood[42], Tomita[35-38], Fahle[13], Regin[29], Konc[21], Kumlander[23], Chu-Min Li[18] and etc.).

The problem of vertex coloring is $\mathcal{NP}$-complete too[19] and therefore MCP algorithms use heuristic vertex coloring techniques(DSATUR, GREEDY and other), for example see [20,28,39] . Such MCP algorithms demonstrate quite good results on the random graphs and more worse results on the special benchmarks, such as DIMACS[11], BHOSLIB[5] or Sloane[33].

Recently appeared a lot of good surveys (Segundo[32], Carmo[8]) in which different MCP algorithms have been compared. Carmo[8] proposed to consider the following:

*"Explaining the gap between the disheartening worst case estimates and what has actually already been achieved in practice seems to be an interesting challenge."*

With this long term goal in mind, we will show that MC algorithms using different heuristic vertex-coloring cannot run better than $\Omega(2^{0.2n})$. Notice that we are concerned with lower bound on the complexity of a particular class of MCP algorithms, and not with lower bounds on the complexity of an Maximum Clique Problem.

## 4. The Join Graph

The *join graph* $G = G_1 + G_2$ of the disjoint graphs $G_1$ and $G_2$ is defined by

$V(G) = V(G_1) \cap V(G_2)$,

$E(G) = E(G_1) \cup E(G_2) \cup \{(v, u) \mid v \in V(G_1),\ u \in V(G_2)\}$

Let's emphasize some properties of the join graph G.

**Proposition 4.1**. Let $n(G_1)$ and $n(G_2)$ are orders of $G_1$ and $G_2$ respectively. Order $n(G)$ of the join graph $G = G_1 + G_2$ is

$$n(G) = n(G_1) + n(G_2).$$

**Proposition 4.2**. Let $\alpha(G_1)$ and $\alpha(G_2)$ are stable numbers of $G_1$ and $G_2$ respectively. Stable number $\alpha(G)$ of the join graph $G = G_1 + G_2$ is

$$\alpha(G) = \max\{\alpha(G_1), \alpha(G_2)\}.$$

**Proposition 4.3**. Let $w(G_1)$ and $w(G_2)$ are clique numbers of $G_1$ and $G_2$ respectively. Clique number $w(G)$ of the join graph $G = G_1 + G_2$ is

$$w(G) = w(G_1) + w(G_2).$$

**Proposition 4.4**. Let $\chi(G_1)$ and $\chi(G_2)$ are chromatic numbers of $G_1$ and $G_2$ respectively. Chromatic number $\chi(G)$ of the join graph $G = G_1 + G_2$ is

$$\chi(G) = \chi(G_1) + \chi(G_2).$$

**Proposition 4.5**. Let $(G_1)$ and $(G_2)$ are chromatic gaps of $G_1$ and $G_2$ respectively. Chromatic gap $\psi(G)$ of the join graph $G = G_1 + G_2$ is

$$\psi(G) = \psi(G_1) + \psi(G_2).$$

## 5. Color-critical graphs

A graph G is called *color-critical* or *vertex-color-critical* or, briefly, *critical* if $\chi(G') < \chi(G)$ for every proper subgraph G' of G. It is called *edge-critical* if $\chi(G - \{e\}) < \chi(G)$ for every edge *e* of G and it is called *k-(edge)-critical* graph if it is (edge-)critical and k-chromatic.

**Proposition 5.1**. (Dirac's construction[16]). *Let $G_1 \in Chr(k_1)$ and $G_2 \in Chr(k_2)$ be disjoint graphs and $G = G_1 + G_2$. Then*

(a) $G \in Chr(k_1 + k_2)$,

(b) $G \in Cri(k_1 + k_2)$ *if and only if* $G_1 \in Cri(k_1)$ *and* $G_2 \in Cri(+ k_2)$.

Let $C_5$ is odd cycle of the length 5. For $C_5$ is straightforward:

$$n(C_5) = 5; \quad \alpha(C_5) = 2; \quad w(C_5) = 2; \quad \chi(C_5) = 3; \quad \psi(C_5) = 1.$$

In 1952 Dirac[16] obtained 6-critical graphs $C_n + C_n$ on $2n$ vertices and $n^2 + 2n$ edges.

We denote such join graph as $C_{n,2}$. Properties of this graph are

$$n(C_{n,2}) = 2n; \quad \alpha(C_{n,2}) = \alpha(C_n); \quad w(C_{n,2}) = 4; \quad \chi(C_{n,2}) = 6; \quad \psi(C_{n,2}) = 2.$$

**Proposition 5.2**. *Let $C_{5,q}$ is the join of q copies of the cycle $C_5$, where $q \geq 2$. So we have 3q-critical graph with properties*

$$n(C_{5,q}) = 5q; \quad \alpha(C_{5,q}) = 2; \quad w(C_{5,q}) = 2q; \quad \chi(C_{5,q}) = 3q; \quad \psi(C_{5,q}) = q.$$

**Proposition 5.3**. *Let $\Gamma_n$ - class of all n-vertex graphs and n is multiple to 5. Then this class contains n-vertex graph $C_{5,q}$, where $q = n/5$, and*

$$\alpha(C_{5,q}) = 2;$$
$$w(C_{5,q}) = 2q = 2n/5 = 0.4n;$$
$$\chi(C_{5,q}) = 3q = 3n/5 = 0.6n.$$
$$\psi(C_{5,q}) = \chi(C_{5,q}) - w(C_{5,q}) = q = n/5 = 0.2n.$$

## 6. The maximum clique algorithm (pseudo-code)

The main idea of maximum clique algorithms which use a heuristic vertex coloring can be described by the below EXACT_MC pseudo-code with a certain simplification of unimportant details. We must note that we do not describe approximate algorithms HEURISTIC_CLIQUE and HEURISTIC_COLORING.

Let us apply procedure START(G) to some graph $G$, which belongs to class $\Gamma_n$. Initially we can apply some vertex set reordering algorithm and then set $level = 0$, $Q_{curr} = \{\phi\}$ and $Q_{max} = \{\phi\}$.

---

*void START(graph G)  {*

    // step A0.  Vertex set reordering
    *REORDERING(V(G));*
    $Q_{max} = \{\phi\}$; // the maximum clique
    $Q_{curr} = \{\phi\}$; // the current clique
    *Calls = 0;*   // number of nodes int branching tree

```
    // step A1. apply some greedy clique algorithm for determine the
    //          lower bound on the order of the maximum clique;
    Q_max = HEURISTIC_CLIQUE(V);
    // call the main step
    EXACT_MC(V(G), 0);
}
```

Simultaneously we focus on the three key issues of the branch and bound method mentioned in the Section 3.

Determine lower bound(step A1): apply some heuristic clique algorithm HEURISTIC_CLIQUE to the input graph $G$. Approximate HEURISTIC_CLIQUE algorithm returns clique Q of the order $|Q|$. So we have $|Q_{max}| = |Q| \geq 1$.

Call EXACT_MC($G$, $0$).

If $|G_{level}|==0$, try to improve solution(step 2) and return.

Determine upper bound(step A3): $|G_{level}| > 0$, apply coloring algorithm HEURISTIC_COLORING and get coloring $C = \{C_1, ..., C_k\}$, where $k=|C|$.

Branching (step A4): While $k=|C| \geq |Q_{max}|$-level, select the last vertex $v$ from the last color class $C_k$, add $v$ into current solution $Q_{curr}$ and build subgraph $G_{level+1}$ equal to the neighborhood $N(v)$ of the vertex $v$. Call EXACT_MC($G_{level+1}$, level + 1).

Because call of EXACT_MC is recursive then than we return(step 7) to the previous *level* we must remove root vertex $v$ from the input subgraph $G_{level}$, from the current solution $Q_{curr}$ and from the last color class $C_k$. If cardinality of $C_k$ equals 0, we reduce $k = k - 1$ and check while-cycle condition(step A4).

```
void EXACT_MC(subgraph G_level, int level) {
    // step A2. Check input subgraph order
    if'(|G_level| == 0) then { // Try to improve solution
        if (|Q_curr| > |Q_max|) then { // improve the solution
            Q_max = Q_curr; |Q_max| = |Q_curr|;
        }
        return;
    }
    ++Calls;
    // step A3. apply some coloring heuristic for determine the upper
    //          bound on the order of the maximum clique;
```

```
            C = HEURISTIC_COLORING(G_level);  // C = {C_1, ..., C_k}, |C|=k;
            k = |C|;
            // step A4. the basic cycle of the algorithm
            while (level + k > | Q_max |) {
                    // step A5. Select root vertex
                    v = the last vertex in the color class C_k;
                    Q_curr = Q_curr + v; // add v into current clique
                    // build next level graph as neigbourhood of the vertex v
                    G_level+1 = N(v, G_level);
                    // step A6. Recursive call
                    EXACT_MC(G_level+1, level + 1);
                    // step A7. Return from the previous level
                    G_level = G_level \ v;         // remove v from the input subgraph;
                    Q_curr = Q_curr \ v; // remove v from the Q_curr
                    -- |C_k|;              // remove v from the color class C_k;
                    if (|C_k|==0)    // check if v was the single vertex in C_k
                            --k;     // reduce number of color classes
            }
            return;
    }
```

**Proposition 6.1**. *Let A – some maximum clique algorithm using vertex-coloring heuristic. The worst case running time of algorithm A is $\Omega(2^{0.2n})$.*

**Proof.** Let $\Gamma_n$ - class of all *n*-vertex graphs and *n* is multiple to 5. Let us note $q = n/5$. Then class $\Gamma_n$ contains graph $C_{5,q}$ which is join of $q$ 5-cycles $C_5$:

$C_{5,q} = C_5^1 + C_5^2 + ... + C_5^q$, where $C_5^i$ is *i*-th cycle $C_5$ and $i = 1, ..., q = n/5$. Let's note cycle $C_5^i = \{ (v_1^i, v_2^i, v_3^i, v_4^i, v_5^i), (e_{12}^i, e_{23}^i, e_{34}^i, e_{45}^i, e_{51}^i) \}$.

Apply algorithm A to the graph $C_{5,q} \in \Gamma_n$.

START, Step A0: graph $C_{5,q}$ vertex set ordering. This step is very important but in this paper we omit the analysis of its efficiency.

START, Step A1 : graph $C_{5,q}$ contains only maximal cliques of order $2q$ **(Proposition 5.2)** then any HEURISTIC_CLIQUE algorithm returns clique $Q(C_{5,q})$ of order $2q$. Set lower bound $|Q_{max}| = |Q(C_{5,q})| = 2q = 2n/5$.

EXACT_MC [ Step A3, *level*=0] : HEURISTIC_COLORING builds coloring $C(C_{5,q})$ of the input subgraph $C_{5,q}$ at least in *k* colors where

$$k = |C| \geq \chi(C_{5,q}) = 3q.$$

In the best case such coloring partitions all vertices into k = *3q* color-classes, where the first *2q* color-classes contain two vertices and remaining *q* classes are singleton(one vertex) color-classes:

| Color-class number | 1 | 2 | 3 | ... | 2q-1 | 2q | 2q+1 | 2q+2 | ... | 3q-1 | 3q |
|---|---|---|---|---|---|---|---|---|---|---|---|
| Color-class members | $v_1^1, v_3^1$ | $v_2^1, v_4^1$ | $v_1^2, v_3^2$ | ... | $v_1^q, v_3^q$ | $v_2^q, v_4^q$ | $v_5^1$ | $v_5^2$ | ... | $v_5^{q-1}$ | $v_5^q$ |

EXACT_MC [Step A4, *level*=0]: this while-cycle will be true for color class numbers from *k = 3q* down to *k ≥ 2q + 1*. So we must repeat this step for *q* times.

EXACT_MC [Step A5, *level*=0]: starting with the vertex $v_5^q$ we build subgraph $G_1$ = $C_{5, q-1}$ + { ($v_1^q, v_4^q$), ( )} with properties *w(G₁) = 2q – 1, $\chi$ (G₁) = 3q – 2* and *ψ(G₁) = q - 1*. The notation { ($v_1^q, v_4^q$), ( )} means 2-vertex edgeless subgraph.

EXACT_MC [ Step A3, *level*=1] : HEURISTIC_COLORING builds coloring C(G₁). In the best case such coloring partitions all vertices of the graph $G_1$ into k = *(3q -2)* color-classes, where the first *2(q – 1) + 1* color-classes contain two vertices and remaining *(q – 1)* classes are singleton(one vertex) color-classes:

| Color-class number | 1 | 2 | 3 | ... | 2q-3 | 2q-2 | 2q-1 | 2q | ... | 3q-3 | 3q-2 |
|---|---|---|---|---|---|---|---|---|---|---|---|
| Color-class members | $v_1^1, v_3^1$ | $v_2^1, v_4^1$ | $v_1^2, v_3^2$ | ... | $v_1^{q-1}, v_3^{q-1}$ | $v_2^{q-1}, v_4^{q-1}$ | $v_2^q, v_4^q$ | $v_5^1$ | ... | $v_5^{q-2}$ | $v_5^{q-1}$ |

EXACT_MC [Step A4, *level*=1]: this while-cycle will be true for color classes number k from *k = 3q - 2* down to *k ≥ 2q*. So we must repeat this step on the *level* 1 for (*q – 1*) times.

EXACT_MC [Step A5, *level*=1]: starting with the vertex $v_5^{q-1}$ we build subgraph $G_2$ = $C_{5, q-2}$ + { ($v_1^q, v_4^q$), ( )} + { ($v_1^{q-1}, v_4^{q-1}$), ( )} with properties *w(G₂) = 2q – 2, $\chi$ (G₂) = 3q – 4* and *ψ(G₂) = q - 2*.

……………………………………………………………..

EXACT_MC [Step A4, *level*=q]:   *q+|C (G_q)|==|Qmax|*, return to the upper *level*.

So existence of the graph $C_{5,q} \in \Gamma_n$ guarantee that an algorithm A cannot run in time better than *Ω(2^{n/5}) = Ω(2^{0.2n})*.

To demonstrate above results we apply algorithm to the join graph C₅, ₅ and describe (Table 6-1) the algorithm behavior.

First of all, we must remember that initially we call START(G)  and build solution lower bound | $Q_{max}$ |= 10 (see Proposition 5.2).

Column 1 of the Table 6-1 contains *level* value. Column 2 displays members of the currently build clique. Triple $G_{level}(n,Q,C)$ in the column 3 describes level input subgraph: subgraph order($n$), it's maximum clique($Q$) and minimal coloring($C$) values. And the last column contains counter of the branching steps.

The first algorithm call corresponds to the *level* = 0, $Q_{curr}$ is empty, input subgraph $G_0 = C_{5,5} = (25, 10, 15)$. In accordance with the step A5 we select the last vertex from the last(=15) color class as the level root vertex $v_0$. In our purposes we set to the level root vertex $v_0$ codename equal to the color class number, i.e. 15. Then we compare *level* + $|C(G_0)| > |Q_{max}|$, add $v_0$ into $Q_{curr}$, build subgraph $G_1 = N(v_0) = (22, 9, 13)$ and go to the next level 1. Here we increase *Calls* value, coloring $G_1$ in 13 colors, check condition *level* + $|C(G_1)| > |Q_{max}|$. In step A5 we select the last vertex from the last(=13) color class as the level root vertex $v_1$, add $v_1$ into $Q_{curr}$, build subgraph $G_2 = N(v_1) = (19, 8, 11)$ and go to the next level 2 and so on up to the *level* 5.

On the *level* 5 we coloring input subgraph $G_5$ into 5 colors and have *level* + $|C(G_5)| == |Q_{max}|$. We return from this level on *level 4*, remove the level root vertex $v_4$ (= 07) from $Q_{curr}$, remove $v_4$ from $G_4$, , remove $v_4$ from $C(G_4)$ and reduce $|C(G_4)|$ up to 6. Then check condition on step A4: *level* + $|C(G_4)| == |Q_{max}|$. We return from this level on *level 3*.

So algorithm performs at least 32 calls, that is for n = 25 we have $\Omega(2^{n/5}) = \Omega(2^5)$.

**Table 6-1    Step by step description of the EXACT_MC algorithm**

| level | $Q_{curr}$(level) | Input subgraph $G_{level}(n,Q,C)$ | Branching step (*Calls*) |
|---|---|---|---|
| 0 | [] | (25,10,15) | 1 |
| 1 | [15] | (22, 9,13) | 2 |
| 2 | [15.13] | (19, 8,11) | 3 |
| 3 | [15.13.11] | (16, 7, 9) | 4 |
| 4 | [15.13.11.09] | (13, 6, 7) | 5 |
| 5 | [15.13.11.09.07] | (10, 5, 5) | 6 |
| 4 | [15.13.11.08] | (12, 6, 6) | 7 |
| 3 | [15.13.10] | (15, 7, 8) | 8 |
| 4 | [15.13.10.08] | (12, 6, 6) | 9 |
| 3 | [15.13.09] | (14, 7, 7) | 10 |
| 2 | [15.12] | (18, 8,10) | 11 |
| 3 | [15.12.10] | (15, 7, 8) | 12 |
| 4 | [15.12.10.08] | (12, 6, 6) | 13 |
| 3 | [15.12.09] | (14, 7, 7) | 14 |
| 2 | [15.11] | (17, 8, 9) | 15 |
| 3 | [15.11.09] | (14, 7, 7) | 16 |
| 2 | [15.10] | (16, 8, 8) | 17 |
| 1 | [14] | (21, 9,12) | 18 |
| 2 | [14.12] | (18, 8,10) | 19 |
| 3 | [14.12.10] | (15, 7, 8) | 20 |
| 4 | [14.12.10.08] | (12, 6, 6) | 21 |
| 3 | [14.12.09] | (14, 7, 7) | 22 |
| 2 | [14.11] | (17, 8, 9) | 23 |
| 3 | [14.11.09] | (14, 7, 7) | 24 |
| 2 | [14.10] | (16, 8, 8) | 25 |
| 1 | [13] | (20, 9,11) | 26 |
| 2 | [13.11] | (17, 8, 9) | 27 |
| 3 | [13.11.09] | (14, 7, 7) | 28 |
| 2 | [13.10] | (16, 8, 8) | 29 |
| 1 | [12] | (19, 9,10) | 30 |
| 2 | [12.10] | (16, 8, 8) | 31 |
| 1 | [11] | (18, 9, 9) | 32 |

## 7. Experimental results

We select some reported exact algorithms for the maximum clique problem

   A. MCC - Wood [42]  + Shell sorting the Largest the First;
   B. iMaxClique – Chu-Min Li[18]  + Bubble sorting the Largest the First;
   C. VC-u – Kumlander[23] + Bubble sorting the Largest the First;
   D. MCQ-Dyn – Konc[21] + dynamic vertex reordering;
   E. MCQ – Tomita [35] + Bubble sorting the Largest the First;
   F. MCR – Tomita[37] ;
   G. MCS – Tomita[38] .

We realize this algorithms on C++ and apply them to the sequence of graphs $C_{5,q}$, $q=1,...,30$. This graphs have order from 5 to 150,  clique number from 2 to 60, chromatic number from 3 to 90 and chromatic gap from 1 to 30,  simultaneously.

In algorithms (A) and  (B) we apply bitwise approach proposed by Wood[42] and some ideas from [1, 22]. Algorithm (C) has been  converted from Visual Basic. Algorithms (C) , (E) – (G) have been implemented like algorithm (D) .

Table 7-1 .    Number of  (nodes * $10^{-3}$) in the branching tree

| C(5,q) graphs | | | Number of solution nodes * 10⁻³ | | | | | | |
|---|---|---|---|---|---|---|---|---|---|
| q | n | Lower bound ($2^{0.2*n}$) *10⁻³ | MCC, Wood, 1996 | MCQ, Tomita, 2003 | VColor-u, Kumlander 2005 | MCQ-Dyn, Konc, 2007 | MCR, Tomita, 2007 | iMaxClique, C.-M. Li, 2010 | MCS, Tomita, 2011 |
| 1 | 5 | 0.002 | 0.003 | 0.004 | 0.003 | 0.003 | 0.004 | 0.004 | 0.004 |
| 2 | 10 | 0.004 | 0.015 | 0.017 | 0.019 | 0.016 | 0.009 | 0.014 | 0.009 |
| 3 | 15 | 0.008 | 0.082 | 0.083 | 0.111 | 0.084 | 0.018 | 0.040 | 0.018 |
| 4 | 20 | 0.016 | 0.478 | 0.483 | 0.653 | 0.277 | 0.035 | 0.114 | 0.035 |
| 5 | 25 | 0.032 | 3 | 3 | 4 | 1 | 0.068 | 0.332 | 0.068 |
| 6 | 30 | 0.064 | 16 | 16 | 23 | 4 | 0.133 | 0.982 | 0.133 |
| 7 | 35 | 0.128 | 97 | 97 | 133 | 10 | 0.262 | 3 | 0.262 |
| 8 | 40 | 0.256 | 570 | 570 | 780 | 36 | 0.519 | 9 | 0.519 |
| 9 | 45 | 0.512 | 3,350 | 3,350 | 4,587 | 112 | 1 | 26 | 1 |
| 10 | 50 | 1 | 19,699 | 19,699 | 26,969 | 319 | 2 | 79 | 2 |
| 11 | 55 | 2 | 115,815 | 115,815 | 158,563 | 1,187 | 4 | 236 | 4 |
| 12 | 60 | 4 | 680,920 | 680,920 | 932,249 | 3,461 | 8 | 709 | 8 |
| 13 | 65 | 8 | 4,003,372 | 4,003,372 | 5,481,027 | 10,536 | 16 | 2,126 | 16 |
| 14 | 70 | 16 | -- | -- | -- | 37,260 | 33 | 6,377 | 33 |
| 15 | 75 | 33 | -- | -- | -- | 107,887 | 66 | 19,132 | 66 |
| 16 | 80 | 66 | -- | -- | -- | 332,359 | 131 | 57,396 | 131 |
| 17 | 85 | 131 | -- | -- | -- | -- | 262 | 172,187 | 262 |
| 18 | 90 | 262 | -- | -- | -- | -- | 524 | -- | 524 |
| 19 | 95 | 524 | -- | -- | -- | -- | 1,049 | -- | 1,049 |
| 20 | 100 | 1,049 | -- | -- | -- | -- | 2,097 | -- | 2,097 |
| 21 | 105 | 2,097 | -- | -- | -- | -- | 4,194 | -- | 4,194 |
| 22 | 110 | 4,194 | -- | -- | -- | -- | 8,389 | -- | 8,389 |
| 23 | 115 | 8,389 | -- | -- | -- | -- | 16,777 | -- | 16,777 |
| 24 | 120 | 16,777 | -- | -- | -- | -- | 33,554 | -- | 33,554 |
| 25 | 125 | 33,554 | -- | -- | -- | -- | 67,109 | -- | 67,109 |
| 26 | 130 | 67,109 | -- | -- | -- | -- | 134,218 | -- | 134,218 |
| 27 | 135 | 134,218 | -- | -- | -- | -- | 268,435 | -- | 268,435 |
| 28 | 140 | 268,435 | -- | -- | -- | -- | -- | -- | -- |
| 29 | 145 | 536,871 | -- | -- | -- | -- | -- | -- | -- |
| 30 | 150 | 1,073,741 | -- | -- | -- | -- | -- | -- | -- |

We apply this algorithms to $C_{5,q}$ join graphs, where $q=1,...,30,$ on the Intel(R) Pentium(R) III Xeon computer (4 CPUs), ~2.5GHz, with 6Gb of memory, Windows Server 2003 Standard x64 Edition and Visual Studio 8. We combined all data into Table 7-1. We use the variable "nodes in the solution branching tree" (or simple "nodes") as measure of search performance. Each algorithm was allowed 3600 seconds, i.e. an hour, of CPU time, and if that limit was exceeded we have a Table 7-1 entry of "—". For convenience of comparison we add additional column "*Lower bound* $= \Omega(2^q)$".

The main results of our experiments are:

number of nodes in the solution branching tree for all considered algorithms require not less than $2^{0.2n}$ nodes;

each of considered algorithms are not able to solve full package $C_{5,q}$, $q=1,...,30$, instances in the allowed (3600 seconds) time; the best results belong to the Tomita MCR and MCS algorithm;

in comparison with DIMACS [16], BHOSLIB[15] and Sloane[17] benchmarks proposed $C_{5,q}$ join graphs collection wasn't solved even for not large graph size, namely for n = 140 vertices.